\documentclass[a4paper]{IEEEtran}
\IEEEoverridecommandlockouts

\usepackage{amsmath,amssymb,amsfonts}
\usepackage{soul}
\usepackage{color}
\usepackage{url}
\usepackage{cite}
\usepackage[detect-all]{siunitx}
\usepackage{tabularx,booktabs}
\usepackage{algorithmic}
\usepackage{graphicx}
\usepackage{pgfplots}
\usepackage{tikzscale}
\usepackage[siunitx]{circuitikz}
\usetikzlibrary{shapes.multipart}
\ifCLASSOPTIONcompsoc
\usepackage[caption=false,font=normalsize,labelfont=sf,textfont=sf]{subfig}
\else
\usepackage[caption=false,font=footnotesize]{subfig}
\fi

\usetikzlibrary{external}
\usepgfplotslibrary{external} 
\usepgfmodule{nonlineartransformations}
\pgfplotsset{compat=1.6}
\usepackage{datatool}
\usepackage{textcomp}
\usepackage{xcolor}
\def\BibTeX{{\rm B\kern-.05em{\sc i\kern-.025em b}\kern-.08em
    T\kern-.1667em\lower.7ex\hbox{E}\kern-.125emX}}

\newcommand{\s}[1]{\mathrm{#1}}
\newcommand{\ph}[3][]{\varphi_\s{#2}^\s{#1}(#3)}  
\renewcommand{\j}{\s{j}}

\let\originalleft\left
\let\originalright\right
\renewcommand{\left}{\mathopen{}\mathclose\bgroup\originalleft}
\renewcommand{\right}{\aftergroup\egroup\originalright}
\DeclareMathOperator{\sinc}{sinc}
\definecolor{myred}{RGB}{255,83,83}
\definecolor{myred2}{RGB}{235, 25, 26}
\definecolor{mygreen}{RGB}{56, 118, 67}

\usepackage{fancyhdr}
\fancypagestyle{empty}{fancy}

\begin{document}

\title{Two-Dimensional Arbitrary Angle of Arrival\\in Radar Target Simulation\\
{\footnotesize} }

\author{\IEEEauthorblockN{Axel Diewald, Benjamin Nuss, Johannes Galinsky and Thomas Zwick}
\\	\IEEEauthorblockA{Karlsruhe Institute of Technology, Karlsruhe, Germany\\
		Email: axel.diewald@kit.edu}
}

\maketitle

\begin{abstract}
Automotive radar sensors play a key role in the current development of advanced driver assistance systems (ADAS). Their ability to detect objects even under adverse weather conditions makes them indispensable for environment-sensing tasks in autonomous vehicles. Since an operational failure presents a potential risk to human life, thorough and practical validation testing must be performed, requiring an integrative test solution. Radar target simulators (RTS) are capable of performing over-the-air validation tests by generating virtual radar echoes that are perceived as targets by the radar under test (RuT). Since the authenticity and credibility of these targets is based on the accuracy with which they are created, their simulated position must be arbitrarily adjustable. In this work, an existing approach to synthesize virtual radar targets at an arbitrary angle of arrival (AoA) is extended to cover both, the azimuth and elevation domain. The concept is based on the superposition of the returning signals from four neighboring RTS channels. A theoretical model describing the basic principle and its constraints is developed. In addition, a measurement campaign is conducted to verify the practical functionality of the proposed approach.
\end{abstract}


\section{Introduction}
\IEEEPARstart{I}{n recent} years the development of advanced driver assistance systems (ADAS) and thus autonomous driving has steadily increased. In order to ensure safe operation of ADAS functions, thorough testing procedures must be implemented, that validate not only the function itself but also the sensing devices they rely on. In addition to other sensors such as camera, lidar, and ultrasound, radar sensors play an important role in the environment sensing tasks of ADAS and must therefore be tested integratively. However, carrying out such validation tests on the road involves a great deal of effort, as distances in the magnitude of hundred million kilometers have to be covered to guarantee that the radar sensors function properly \cite{MGLW2015,S2017,KW2016}. Furthermore, these tests are unrepeatable, as individual traffic situations can not be reproduced identically.

For these reasons, radar target simulators (RTS) have recently attracted much attention in research and commercially because of their ability to thoroughly validate radar sensors in-place and under laboratory conditions \cite{GSGVABMPP2018, IRW2020, WMNLD2020}. Their operating principle is to deceive a radar under test (RuT) by generating an artificial environment of virtual radar targets. To make this environment as credible and realistic as possible, the virtual radar targets must be generated as accurately as possible in terms of their properties. Recent RTS systems have already achieved a set-point precision that is higher than the resolution of conventional and even future radars in regards of range, Doppler and radar cross section (RCS) \cite{GMS18, SD2018, 9371304, 9448793}.

However, the simulation of the angle of arrival (AoA) in both azimuth and elevation is not yet able to achieve the angle estimation capabilities of their counterpart. So far, there have been approaches to electronically switch between discrete and fixed angular positions to simulate the azimuth or elevation dislocation of a virtual radar target. Regardless of whether they implement only one \cite{GMS18,RS2021} or both angular dimensions \cite{KT2021} at the same time, their ability to simulate angles lags behind the angular accuracy capabilities of modern radar sensors. Another approach is to mechanically rotate the RTS system centric around the RuT \cite{GR2017,KC20xx} in addition to moving the RTS vertically \cite{9337461}, which significantly limits the number of virtual targets and their inherent lateral movement velocity. In \cite{KT2017} the RuT itself is rotated, which results in the same restrictions and, in addition, is not suitable for integrated radar sensor tests.

\begin{figure}[!t]
	\centering
	\input{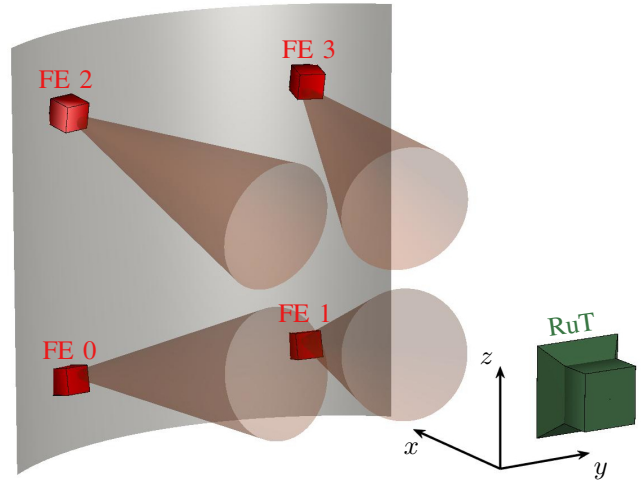}
	\caption{Concept of the system setup.}
	\label{fig_rts_concept}
\end{figure}

Therefore, the authors present a new approach that allows virtual radar targets to be generated with an arbitrary AoA in both azimuth and elevation simultaneously. The basic concept was already presented in \cite{9525811}, but was limited to the simulation of azimuth angles and has now been extended. The approach is not limited in terms of the RTS's methodology, as it is applicable to analog and digital systems, nor is it limited by the modulation scheme of the RuT. It is based on the superposition of four adjacent virtual radar echoes, that enables the synthesis of simulated radar targets at an arbitrarily adjustable lateral position. For this purpose, the four associated RTS channels are arranged in a square formation. The concept of the system setup is shown in Fig. \ref{fig_rts_concept}. 

In the following the basic operation of radar target simulation, the underlying signal model and the functional principle of the new approach will be explained. Thereupon, a measurement campaign, that includes calibration measurements and a demonstration of the functioning of the proposed method, is presented.

\section{Radar Target Simulation}

Since detailed descriptions of the general working principle of RTS systems can be found extensively in the literature \cite{GMS18,DMW18,s20092714}, only their fundamental operation shall be explained here. An RuT is placed closely in front of the RTS antenna front ends (FE). The signal transmitted by the radar is received by the FEs and down converted to a lower intermediate frequency $f_\s{rts}$. Subsequently, the target generation modifications are applied to the signal before it is up converted back to its original carrier frequency $f_\mathrm{c}$ and re-transmitted by the FEs towards the RuT. The signal modifications, namely time delay, Doppler shift and attenuation, that enable the virtual radar target generation can be implemented in either the analog \cite{EPB2016,LEWW2014,GGSABMP2017} or digital domain \cite{WMNLD2020,9396154,9371304}. Since the concept presented here is not limited in terms of the RTS's target generation methodology, the subsequent analytical elaborations are kept generic.


\subsection{Signal Model}

In the following the radar signal that is transmitted by the RuT and modified by the RTS will be modeled in order to present the relation between the signal amplitude and the detected AoA. As will later be shown, the signal phase plays an important role for the success of the proposed method. Therefore, the following analytical descriptions focus primarily on the signal phase in order to facilitate the comprehension of the approach.

For the sake of simplicity, a frequency-modulated continuous wave (FMCW) radar will be assumed, whose signal frequency and phase can be described in regards of time $t \in [0,T]$ as

\begin{align}
	f_\s{tx}(t) &= f_\s{c} + \frac{B}{T} \cdot t \\
	\ph{tx}{t} &= 2\pi \int_0^{t} f(t')\cdot dt' = 2\pi \left[f_\s{c} \cdot t + \frac{B}{2T} \cdot t^2\right]
\end{align}
where $B$ describes the signal's bandwidth and $T$ the chirp period. After its travel through free space and its reception, modification and re-transmission by the RTS, the signal is received by the RuT and mixed with the original transmit signal to form the beat signal. Thereupon, it is discretized by the radar's analog-to-digital converter (ADC) and a discrete Fourier transform (DFT) and subsequent detection is applied. Thereafter, the signal can be expressed as 
\begin{align}
	x_\s{R}[f_\s{R}] =& \ A_q N_\s{s} \cdot \exp \left\{ \j \varphi_\s{R}[f_\s{R}] \right\} \label{eq_x_range} \\
	\varphi_\s{R}[f_\s{R}] =& \ 2\pi \left[f_\s{c} \tau_\s{c} + f_\s{rts} \tau_\s{rts} + \frac{1}{2} \left( B \tau -  f_\s{R} \right)\right]
\end{align}
where $f_\s{R}$ is the DFT bin index, $A_q$ the RTS's attenuation, $N_\s{s}$ the number of samples, $\tau_\s{c}$ the free space propagation delay, $\tau_\s{rts}$ the RTS's delay and $\tau = \tau_\s{c} + \tau_\s{rts}$ the total time delay between the receive and transmit signal. The free space propagation delay ($\tau_\s{c} = \tau_\s{tx} + \tau_\s{rx}$) combines the delay of the transmitted and returning signal. A more detailed derivation of the intermediate steps can be found in \cite{9525811}.

Taking into account the RuT's multiple-input multiple-output (MIMO) antenna array, comprising of $N_\s{tx}$ transmit and $N_\s{rx}$ receive antenna elements, the returning signal's free space propagation delay
\begin{align}
	\tau_\s{rx} &= \frac{ R_\s{c} + y \cdot \sin(\theta_q)\cos(\psi_q) + z \cdot \sin \left(\psi_q\right)}{c_\s{0}}
\end{align}
is dependent on the physical distance between the RuT and the RTS $R_\s{c}$ and the azimuth and elevation angle $\theta_q, \psi_q \in [\SI{-90}{\degree},\SI{90}{\degree}]$ at which the respective RTS FE ($q \in [0,\ldots,Q-1]$) is located as seen by the RuT. It is also conditional on the horizontal and vertical positions of the antenna elements
\begin{align}
	y &= d_\s{y,tx} \cdot n_\s{tx} + d_\s{y,rx} \cdot n_\s{rx} \\
	z &= d_\s{z,tx} \cdot n_\s{tx} + d_\s{z,rx} \cdot n_\s{rx}
\end{align}
where $d_\s{y/z,tx/rx}$ denotes the horizontal\allowbreak/\allowbreak vertical transmit\allowbreak/\allowbreak receive antenna element spacing and $n_\s{tx/rx} \in\allowbreak[0,\ldots,N_\s{tx/rx}-1]$ indexes the antenna elements.
Applying beamforming allows to estimate the AoA
\begin{align}
	x_{\s{A},q}[\alpha,\beta] =& \sum_{n_\s{tx}=0}^{N_\s{tx}-1} \sum_{n_\s{rx}=0}^{N_\s{rx}-1} x_{\s{R},q} [f_\s{R}] \nonumber\\
	& \exp \left\{-\j2\pi\frac{y \cdot \sin \left(\alpha\right) \cos \left(\beta\right) + z \cdot \sin \left(\beta\right)}{\lambda} \right\}
\end{align}
where $\lambda$ is the radar signal's wave length and $\alpha,\beta \in [\SI{-90}{\degree},\SI{90}{\degree}]$ are orientated equal to $\theta_q$ and $\psi_q$. The expression can be simplified using the partial sum of a geometric series \cite{Bron01}, $\cos(x) \approx 1$ and $\sin(x) \approx x$ for $|x| \ll 1 $. In addition, common radar sensors have a one-dimensional distribution of their transmit and receive antennas, which allows e.g. $d_\s{y,tx}$ and $d_\s{z,rx}$ to be set to zero, leading to
\begin{align}
	x_{\s{A},q}[\alpha,\beta] =& \; A_q N_\s{s} N_\s{tx} N_\s{rx} \cdot \exp \left\{ \j \varphi_{\s{A},q} \right\} \nonumber\\
	&  \sinc \left( \frac{N_\s{tx}}{\lambda} \left(\sin(\psi_q) - \sin(\beta)\right) d_\s{z,tx} \right)  \nonumber\\
	& \sinc \left( \frac{N_\s{rx}}{\lambda} \left(\sin(\theta_q) - \sin(\alpha)\right) d_\s{y,rx} \right) \\
	\varphi_{\s{A},q} =& \; 2\pi  \left[ \left( f_\s{c} + \frac{B}{2} \right) \frac{2 R_\s{c}}{c_\s{0}} + \left( f_\s{rts} + \frac{B}{2} \right) \tau_\s{rts} \right. \nonumber\\
	& \qquad + \sin(\theta_q) \cdot  \frac{d_\s{y,rx}}{2\lambda} \cdot (N_\s{rx}-1) \nonumber\\
	& \qquad \, \left. + \sin(\psi_q) \cdot \frac{d_\s{z,tx}}{2\lambda} \cdot (N_\s{tx}-1) \right]
\end{align}
The $\s{sinc}$-function bares a maximum, where its argument equals zero, resulting in a target being detected at $\sin(\alpha) = \sin(\theta_q)$ and $\sin(\beta) = \sin(\psi_q)$.

\subsection{Superposition}

The basic principle of the approach presented here lies in the superposition of the returning signals of four ($Q = 4$) RTS channels that are arranged in a squared formation. The detected AoA is situated in between the physical positions of the respective FEs and can be controlled by means of their attenuation. In order for the beat signals to successfully overlay additively, the individual phases must be coherent
\begin{align}
	\varphi_{\s{A},q} = \varphi_{\s{A}} \qquad \forall \qquad q \in [0,\ldots,Q-1]
\end{align}
Furthermore, due to the arrangement of the FEs, the angle variables can be partially equated to form a left ($\theta_\s{l} = \theta_\s{0} = \theta_\s{2}$), right ($\theta_\s{r} = \theta_\s{1} = \theta_\s{3}$), bottom ($\psi_\s{b} = \psi_\s{0} = \psi_\s{1}$) and top ($\psi_\s{t} = \psi_\s{2} = \psi_\s{3}$) angular position. Subsequently, the signal amplitudes arising from the RTS's attenuation can be expressed as follows
\begin{align}
	A_\s{0} &= A_\s{b} \cdot A_\s{l} \qquad A_\s{1} = A_\s{b} \cdot A_\s{r} \nonumber\\
	A_\s{2} &= A_\s{t} \cdot A_\s{l} \qquad A_\s{3} = A_\s{t} \cdot A_\s{r} 
\end{align}
The resulting superimposed signal can be described as
\begin{align}
	\widehat{x}_\s{A}(\alpha,\beta) =& \sum_{q=0}^{Q-1} x_{\s{A},q}(\alpha,\beta) \nonumber\\
	=& N_\s{s} N_\s{tx} N_\s{rx} \cdot \exp \left\{ \j \varphi_{\s{A}} \right\} \nonumber\\
	& \left[ A_\s{l} \sinc \left( \frac{N_\s{rx}}{\lambda} \left(\sin(\theta_\s{l}) - \sin(\alpha)\right) d_\s{y,rx} \right) \right. \nonumber\\
	& \left. + A_\s{r} \sinc \left( \frac{N_\s{rx}}{\lambda} \left(\sin(\theta_\s{r}) - \sin(\alpha)\right) d_\s{y,rx} \right) \right] \nonumber\\
	&\left[ A_\s{b} \sinc \left( \frac{N_\s{tx}}{\lambda} \left(\sin(\psi_\s{b}) - \sin(\beta)\right) d_\s{z,tx} \right) \right. \nonumber\\
	& \left. + A_\s{t} \sinc \left( \frac{N_\s{tx}}{\lambda} \left(\sin(\psi_\s{t}) - \sin(\beta)\right) d_\s{z,tx} \right) \right]
\end{align}
For the sake of clarity, the following substitution will be employed
\begin{align}
	g_q[\alpha] =& \; \sinc \left( \frac{N_\s{rx}}{\lambda} \left(\sin(\theta_q) - \sin(\alpha)\right) d_\s{y,rx} \right) \\
	g_q[\beta] =& \; \sinc \left( \frac{N_\s{tx}}{\lambda} \left(\sin(\psi_q) - \sin(\beta)\right) d_\s{z,tx} \right)
\end{align}
To identify how the respective RTS channel attenuation affects the location of the maximum of the superimposed signal and therefore the detected AoA, the partial derivations according to $\alpha$ and $\beta$ are formed and set to zero
\begin{align}
	\frac{\partial\widehat{x}_\s{A}(\alpha,\beta)}{\partial\alpha} =& N_\s{s} N_\s{tx} N_\s{rx} \cdot \exp \left\{ \j \varphi_{\s{A}} \right\} \cdot \left( A_\s{b} g_\s{0}[\beta] + A_\s{t} g_\s{1}[\beta] \right) \nonumber\\
	& \left( A_\s{l} \frac{\partial g_\s{0}[\alpha]}{\partial\alpha} + A_\s{r} \frac{\partial g_\s{1}[\alpha]}{\partial\alpha} \right) = 0 \label{eq_atten_a}\\
	\frac{\partial\widehat{x}_\s{A}(\alpha,\beta)}{\partial\beta} =& N_\s{s} N_\s{tx} N_\s{rx} \cdot \exp \left\{ \j \varphi_{\s{A}} \right\} \cdot \left( A_\s{l} g_\s{0}[\alpha] + A_\s{r} g_\s{1}[\alpha] \right) \nonumber\\
	& \left( A_\s{b} \frac{\partial g_\s{0}[\beta]}{\partial\beta} + A_\s{t} \frac{\partial g_\s{1}[\beta]}{\partial\beta} \right) = 0 \label{eq_atten_b}
\end{align}
Considering the basic principle, that a product equals zero if one of its factors does, the derived part of the expressions above can be extracted to calculate the required attenuations for a specific AoA. For reasons of completeness, the derivatives of the substitution are given as
\begin{align}
	\frac{\partial g_q[\alpha]}{\partial\alpha} =& \frac{\cos(\alpha) \sin\left( \pi \frac{N_\s{rx}}{\lambda} \left(\sin(\theta_q) - \sin(\alpha)\right) \cdot d_\s{y,rx} \right)}{\pi \frac{N_\s{rx}}{\lambda} d_\s{y,rx} (\sin(\theta_q) - \sin(\alpha))^2} \nonumber \\
	& - \frac{\cos(\alpha) \cos\left( \pi \frac{N_\s{rx}}{\lambda} (\sin(\theta_q) - \sin(\alpha)) d_\s{y,rx} \right)}{\sin(\theta_q) - \sin(\alpha)}\\
	\frac{\partial g_q[\beta]}{\partial\beta} =& \frac{\cos(\beta) \sin\left( \pi \frac{N_\s{tx}}{\lambda} \left(\sin(\psi_q) - \sin(\beta)\right) d_\s{z,tx} \right)}{\pi \frac{N_\s{tx}}{\lambda} d_\s{z,tx} (\sin(\psi_q) - \sin(\beta))^2} \nonumber \\
	& - \frac{\cos(\beta) \cos\left( \pi \frac{N_\s{tx}}{\lambda} (\sin(\psi_q) - \sin(\beta)) d_\s{z,tx} \right)}{\sin(\psi_q) - \sin(\beta)}
\end{align}


\section{Measurement}

\begin{figure}[!t]
	\centering
	\input{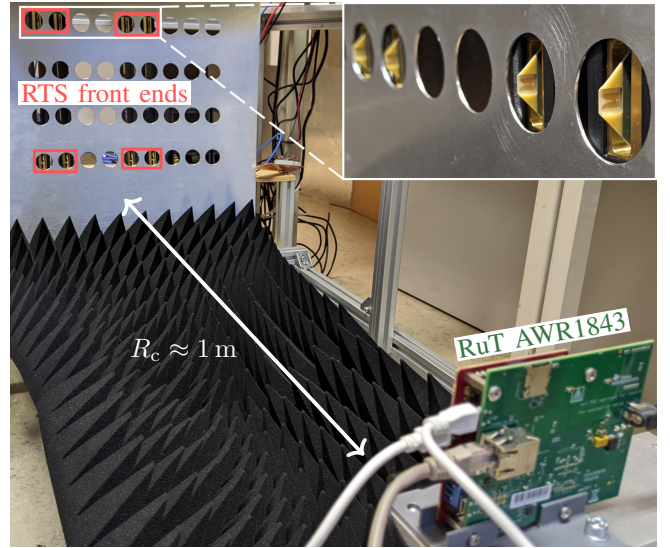}
	\caption{Photograph of the measurement setup.}
	\label{fig_meas_setup}
\end{figure}

\begin{table}[!t]
	\renewcommand{\arraystretch}{1.25}
	\caption{RTS Front End Angular Positions}
	\label{tab_fe_pos}
	\centering
	\begin{tabular}{crrr}
	\toprule
	Front end & Azimuth & Elevation\\
	\midrule
	0 & $\SI{-5.4}{\degree}$ & $\SI{-8.8}{\degree}$ \\
	1 & $\SI{4.5}{\degree}$ & $\SI{-7.7}{\degree}$ \\
	2 & $\SI{-3.4}{\degree}$ & $\SI{8.4}{\degree}$ \\
	3 & $\SI{3.8}{\degree}$ & $\SI{9.9}{\degree}$ \\
	\bottomrule
\end{tabular}

\end{table}

A measurement was conducted utilizing a digital RTS system whose individual components are presented in \cite{9448793} and \cite{vehicles3020016}. The measurement setup is depicted in Fig. \ref{fig_meas_setup} where it can been seen, that the FEs were positioned in a semicircular formation in the horizontal plane with a radius of $R_\s{c} \approx \SI{1}{\meter}$ and in a squared formation in the vertical plane. The respective angular positions of the FEs are given in Table \ref{tab_fe_pos}. A metal sheet facilitated the placement of the FEs and shielded off most of the unwanted static radar reflections of the background. The back end of the RTS was realized with an UltraScale+ RFSoC FPGA from Xilinx. The RTS's intermediate frequency was set to $f_\s{rts} = \SI{500}{\mega\hertz}$. For the RuT a Texas Instruments AWR1843BOOST radar board was employed. It was configured to use all $N_\s{tx} = 3$ transmit and $N_\s{rx} = 4$ receive antennas, a bandwidth of $B = \SI{1}{\giga\hertz}$ and a carrier frequency of $f_\s{c} = \SI{77}{\giga\hertz}$.


\begin{figure}[!t]
	\centering
	\includegraphics[width=1\columnwidth]{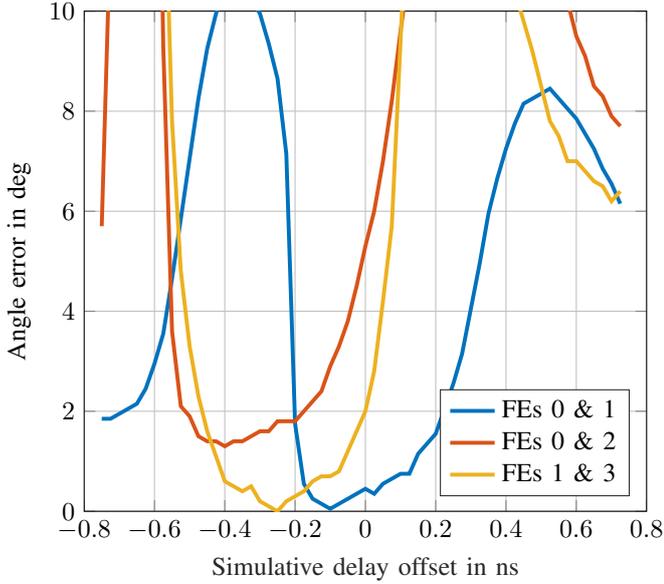}
	\caption{Measured angle error over artificial delay offset used for phase coherency calibration of the respective RTS front ends.}
	\label{fig_meas_ang_calib}
\end{figure}

First, a calibration measurement to create phase coherency among all RTS channels, according to the detailed explanations in \cite{9525811}, was deducted, as it is a prerequisite for a successful additive superposition of the individual echo signals. The calibration was performed stepwise with only two FEs active at a time. The delay of one of the channels was kept constant, and the respective angle error ($\alpha_\s{\epsilon}$ or $\beta_\s{\epsilon}$) of the superimposed target was monitored while sweeping the delay of the other channel. Fig. \ref{fig_meas_ang_calib} depicts the individual calibration procedures. The observable angle error minima were utilized to set the RTS channels' delay offsets for the succeeding measurements.

\begin{figure}[!t]
	\centering
	\includegraphics[width=1\columnwidth]{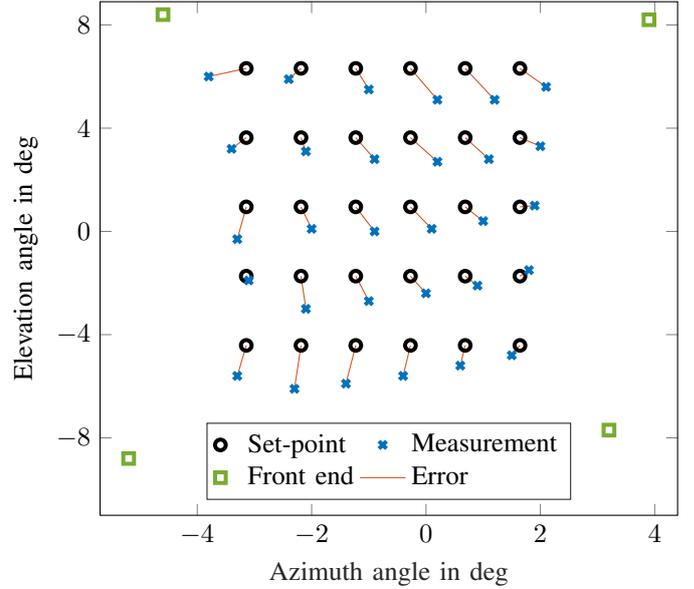}
	\caption{Two-dimensional arbitrary AoA measurement.}
	\label{fig_meas_sim_ang}
\end{figure}

For the two-dimensional virtual target angle simulation, 30 individual measurements, each with a single superimposed target, were conducted. The target position was varied between six azimuth and five elevation angles and the corresponding RTS channel attenuations were determined according to \eqref{eq_atten_a} and \eqref{eq_atten_b}. Fig. \ref{fig_meas_sim_ang} shows the set and measured angular position, as well as their respective deviation of all measurements. It can be assumed that the angle errors occur due to the imperfect FE alignment and a remaining amplitude and phase offset between the RTS channels. In particular, the phase calibration presents itself as a complex task, since phase coherency can only be achieved metrologically in pairs, but is required between all channels simultaneously.

%
%
\begin{figure}
	\centering
	\subfloat[Simulated and measured azimuth angle error.\label{meas_sim_ang_err_az}]{%
		\includegraphics{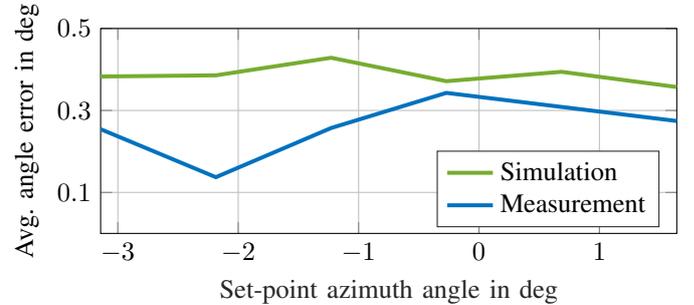}}\\
	\subfloat[Simulated and measured elevation angle error.\label{meas_sim_ang_err_el}]{%
		\includegraphics{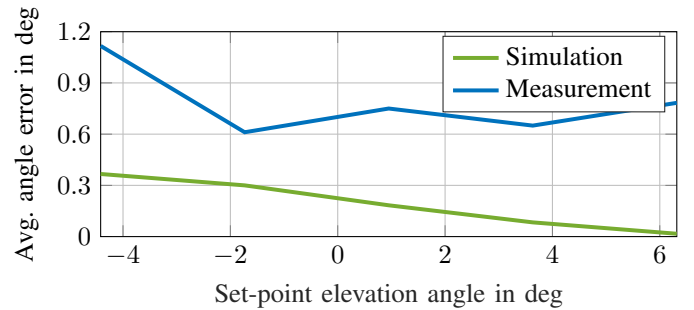}}\\
	\caption{Simulated and measured (a) azimuth and (b)~elevation angle error.}
	\label{meas_sim_ang_err} 
\end{figure}

In Fig. \ref{meas_sim_ang_err} the resulting average azimuth and elevation angle errors are shown as a function of the nominal angles. The simulated values serve as a reference of the expected error due to the inaccuracies in the FE placement. Their inaccurate horizontal and vertical alignment contradicts the simplification constraints in the amplitude calculation, which are, however, necessary to enable an analytically unambiguous solution.


\section{Conclusion}

The proposed approach extends a previously developed concept to allow radar target simulators to generate virtual targets at an arbitrary angle of arrival in both the azimuth and elevation domain. Mathematical analysis of the signal model presented reveals the constraints that have to be met for a successful control of the simulated angle. A calibration method to fulfill these requirements was developed and performed. The approach was implemented on a digital radar target simulator and the measurement campaign conducted verifies the practical functionality of the two-dimensional arbitrarily adjustable AoA.

\section*{Acknowledgment}

The authors would like to thank PKTEC GmbH, Schutterwald, Germany, for providing the front-end transceiver hardware and Texas Instruments Inc., Dallas, TX, USA, for supplying the radar under test (RuT).

\bibliographystyle{IEEEtran}
\bibliography{literature.bib}

\end{document}